\documentclass[aps,prd,floats,nofootinbib,preprintnumbers,superscriptaddress]{revtex4}
\usepackage{epsfig}

\begin{document}

\preprint{NUHEP-TH/05-07}

\title{Non-Oscillation Probes of the Neutrino Mass Hierarchy and Vanishing $|U_{e3}|$}

\author{Andr\'e de Gouv\^ea}
\affiliation{Northwestern University, Department of Physics \& Astronomy, 2145 Sheridan Road, Evanston, IL~60208, USA}

\author{James Jenkins}
\affiliation{Northwestern University, Department of Physics \& Astronomy, 2145 Sheridan Road, Evanston, IL~60208, USA}


\begin{abstract}
One of the outstanding issues in neutrino physics is the experimental determination of the neutrino mass hierarchy: Is the order of the neutrino masses ``normal'' --- $m_1^2<m_2^2<m_3^2$ --- or is it inverted --- $m_3^2<m_1^2<m_2^2$, with $m_2^2-m_1^2\ll m_2^2,m_1^2$? While this issue can be resolved in next-generation long-baseline $\nu_{\mu}\leftrightarrow\nu_e$ neutrino oscillation studies if $|U_{e3}|^2\equiv\sin^2\theta_{13}$ is large enough, a clear strategy on how to resolve it if $|U_{e3}|^2$ is sufficiently small is still lacking. 

Here, we study the capability of non-oscillation probes of neutrino masses to determine the neutrino mass ordering. We concentrate on studies of $m_{\nu_e}$, the kinematical neutrino mass to which precise studies of tritium $\beta$-decay are sensitive, $m_{ee}$, the effective mass to which the rate for neutrinoless double-beta decay is sensitive if the neutrinos are Majorana fermions, and $\Sigma$, the sum of the neutrino masses, to which cosmological probes of the energy budget of the Universe are sensitive. We find that combined measurements of $m_{ee}$, $\Sigma$, and $m_{\nu_e}$ are capable of establishing the neutrino mass hierarchy if these measurements are precise enough and if one ``gets lucky.'' We quantify the previous sentence in detail by performing a numerical analysis of a large number of theoretical data sets, for different measured values of $m_{\nu_e}$, $m_{ee}$, and $\Sigma$, keeping in mind the ultimate sensitivity that can be reached by next (and next-to-next) generation experiments.

\end{abstract}

\maketitle

\setcounter{equation}{0}
\setcounter{footnote}{0}

\section{Introduction}
\label{sec:intro}

In order to explain all neutrino data \cite{TASI} (with the exception of those from LSND \cite{LSND}, waiting to be confirmed by the ongoing MiniBooNE experiment), one is required to augment the standard model Lagrangian by adding operators that will render, after electroweak symmetry breaking, the neutrinos massive \cite{neutrino_theory}.

Under the most conservative assumptions, the new standard model parameters that need to be determined from experiment are:
\begin{enumerate}
\item The three neutrino masses, $m_i$, $i=1,2,3$, chosen real and nonnegative. These are ordered as follows: $m_1^2<m_2^2$, and $0<\Delta m^2_{12}\equiv m^2_2-m^2_1<|\Delta m^2_{13}|$, where $\Delta m^2_{13}\equiv m^2_3-m^2_1$. A positive value of $\Delta m^2_{13}$ implies $m_3^2>m_2^2$ and a so-called normal mass hierarchy, while a negative value of $\Delta m^2_{13}$ implies $m_3^2<m_1^2$ and a so-called inverted mass hierarchy.

The three neutrino masses can be expressed in terms of $\Delta m^2_{12}$, $\Delta m^2_{13}$ (including its sign) and the lightest neutrino mass $m_l$. In the case of a normal mass hierarchy, $m_1=m_l$, $m_2=\sqrt{\Delta m^2_{12}+m^2_l}$, and $m_3=\sqrt{\Delta m^2_{13}+m^2_l}$, while if the mass hierarchy is inverted $m_1=\sqrt{-\Delta m^2_{13}+m^2_l}$, $m_2=\sqrt{-\Delta m^2_{13}+\Delta m^2_{12}+m^2_l}$, and $m_3=m_l$.

\item The six mixing parameters that characterize the leptonic mixing matrix. We adopt the particle data group parameterization of the leptonic mixing matrix (Eq. (13.32) in \cite{pdg}). Two of the six parameters are unphysical if the neutrinos are Dirac fermions.
\end{enumerate}

Current experiments have been able to measure $\Delta m^2_{12}$ and $|\Delta m^2_{13}|$, together with the ``solar angle'' $\theta_{12}$ and the ``atmospheric angle'' $\theta_{23}$. The third mixing angle $\theta_{13}$ is constrained to be small ($\sin^2\theta_{13}<0.03$ at the 99\% confidence level \cite{global_anal}), and we know nothing about the three CP-odd phases $\delta, \alpha_1, \alpha_2$. We also don't know whether the neutrino mass hierarchy is normal or inverted, and the lightest neutrino mass is only modestly constrained: $m_l\in[0,2.0]$~eV, at the 99\% confidence level \cite{global_anal}.\footnote{This is the most conservative bound on $m_l$, provided by precision measurements of the end point spectrum of tritium beta-decay \cite{pdg}, and ignores stronger constraints from searches for neutrinoless double-beta decay (which apply only if the neutrinos are Majorana fermions) and from cosmology (which rely on several nontrivial assumptions regarding the evolution of the Universe and its particle content). We will discuss these in more detail later.}
It is among the main goals of fundamental physics experiments to determine the currently unknown leptonic parameters.

In \cite{ours}, together with Boris Kayser, we explored the issue of determining the mass hierarchy with neutrino oscillation experiments. In particular, we discussed the challenges associated with determining the neutrino mass hierarchy in the advent that $\sin^2\theta_{13}$ is vanishingly small. We concluded that, if $\theta_{13}$ happens to be small enough, the only way to determine the neutrino mass hierarchy with neutrino oscillation experiments would be to probe $\nu_{\mu}\to\nu_{\mu}$ and/or $\bar{\nu}_{\mu}\to\bar{\nu}_{\mu}$ oscillations at very long baselines ($L\gtrsim 5000$~km) and relatively small energies ($E_{\nu}\lesssim 500$~MeV). Given the enormous experimental challenges which one needs to overcome in order to achieve the conditions outlined in \cite{ours}, it is of the utmost importance to explore other probes of the neutrino mass hierarchy.

Here, we study non-oscillation probes of neutrino masses and mixing, and determine their capabilities for extracting the neutrino mass hierarchy. We concentrate on three distinct observables: (i) the electron-type neutrino effective mass $m_{\nu_e}$, which can be probed in precise measurements of the end point of nuclear $\beta$ decay spectra, (ii) the effective neutrino mass $m_{ee}$ that characterizes the rate of neutrinoless double double beta decay (if the neutrinos are Majorana particles), and (iii) the sum of the active neutrino masses $\Sigma$, which is constrained, within the concordance cosmological model, by measurements of ``cosmological'' observables. All three are described in detail in Sec.~\ref{sec:probes}. We spell out the circumstances ({\it e.g.}, how precisely should $m_{\nu_e}$, $m_{ee}$, and $\Sigma$ be  constrained) under which the observables listed above can add to our understanding of the neutrino mass spectrum in Sec.~\ref{sec:fits}, emphasizing the importance of combining all three measurements.

We will concentrate on the limit $\sin^2\theta_{13}\to 0$, when ``standard'' matter affected $\nu_{\mu}\leftrightarrow\nu_e$ searches (see \cite{ours} and references therein) for the mass hierarchy are known to fail. Under these circumstances, as emphasized above and in \cite{ours}, alternative probes of the character of the neutrino mass are not only welcome but absolutely necessary.

\setcounter{footnote}{0}
\setcounter{equation}{0}
\section{Non-Oscillation Probes of the Neutrino Mass}
\label{sec:probes}

The only evidence for neutrino masses comes from long-baseline, neutrino oscillation experiments. These take advantage of quantum mechanical interference in order to be sensitive to very small neutrino mass-squared differences.\footnote{Hence the need for very long baselines --- $L_{\rm osc}\propto E_{\nu}/\Delta m^2$ \cite{TASI}. It is always amusing to remember that solar neutrino experiments are sensitive to $\Delta m^2\gtrsim 10^{-11}$~eV$^2$.} On the other hand, neutrino oscillations are impotent when it comes to determining the neutrino masses themselves --- they are only sensitive to the differences among the various neutrino masses.

Of course, several other phenomena are also potentially sensitive to nonzero neutrino masses. We discuss three qualitatively distinct ``non-oscillation probes'' of nonzero neutrino masses, and explore their potential for determining the nature of the neutrino mass hierarchy under the assumption that it will not be revealed by future oscillation experiments.

\subsection{$m_{\nu_e}$ --- Energy Spectrum of Nuclear $\beta$-Decay}

Conservation of energy and momentum dictates that any physical process involving neutrino absorption, emission, or scattering is sensitive to the values of the neutrino masses. Due to the smallness of neutrino masses, however, such effects are, in practice, almost always hopelessly unobservable.

Some of the most promising probes are decay processes in which the neutrino energy and momentum are well constrained and ``easy'' to measure with great precision and/or as small as possible. Among these are multi-body final state $\tau$ decays, $\pi^+$ decay at rest, and nuclear beta-decay, which are said to constrain, respectively, the tau-type, muon-type, and electron-type neutrino effective masses, which will be properly defined below.

Given the current constraints on neutrino masses-squared and neutrino mixing, however, the most precise probe, by far, of kinematical neutrino mass effects is the precise determination of the end point of the $\beta$-ray energy spectrum of tritium decay. The `end point' is defined as the maximum energy that the daughter $\beta$-ray is kinematically allowed to have.

In a little more detail, the $\beta$-ray spectrum is a function of the three neutrino masses-squared $m_i^2$ and can be written, very schematically, as
\begin{equation}
K=|U_{e1}|^2F(m_1^2/E_{\nu}^2,E_{\nu})+|U_{e2}|^2F(m_2^2/E^2_{\nu},E_{\nu})+|U_{e3}|^2F(m_3^2/E_{\nu}^2,E_{\nu}),
\label{eq:mb}
\end{equation}
where $K$ is the $\beta$-ray spectrum as a function of the neutrino energy $E_{\nu}$,\footnote{The neutrino energy is trivially related to the energy of the $\beta$-ray, which is experimentally accessible.} and $F$ is a function of the neutrino energy and the neutrino mass-squared in units of the neutrino energy. In order to obtain Eq.~(\ref{eq:mb}), we assume that tritium beta decay is described by $^3{\rm H}\to ^3{\rm He}+e^-+\bar{\nu}_i$, and that the probability that a specific mass eigenstate $\nu_i$ is emitted is $|U_{ei}|^2$.

Eq.~(\ref{eq:mb}) is independently sensitive to all three neutrino masses as long as $|U_{ei}|^2$ are known (which is, to a good approximation, the case) \cite{farzan_smirnov}. In the case of very small neutrino masses ($m_i^2/E_{\nu}^2\ll 1$), however, this is not useful in practice. We can write $K$, after expanding $F$ around $m_i^2/E_{\nu}^2=0$, as
\begin{eqnarray}
K&=&\sum_i|U_{ei}|^2\left(F_0+\frac{m_i^2}{E_{\nu}^2}F'_0\right)+O\left(\frac{m_i^4}{E_{i}^4}\right), \\
&=&F_0+\frac{m_{\nu_e}^2}{E_{\nu}^2}F'_0+O\left(\frac{m_i^4}{E_{i}^4}\right),
\end{eqnarray}
where $F_0=F$ evaluated at $m_i^2=0$, $F'_0=\partial F/\partial(m^2_i/E^2_{\nu})$ evaluated at $m_i^2=0$, and
\begin{equation}
m_{\nu_e}^2\equiv\sum_i|U_{ei}|^2m_i^2
\label{eq:mb_def}
\end{equation}
is defined as the electron neutrino effective mass-squared. Note that we made use of the fact that $\sum_i |U_{ei}|^2=1$. Henceforth, we will assume, given current bounds on $m_l$ and the neutrino oscillation parameters, that tritium beta-decay experiments are sensitive only to $m^2_{\nu_e}$.

Experimentally, $m_{\nu_e}^2$ is currently constrained to be less than $(2.0)^2$~eV$^2$ at the 99\% confidence level \cite{pdg,global_anal}. Given that this upper bound is much larger than  $\Delta m^2_{12}$ and $|\Delta m^2_{13}|$, the current bound on $m_l^2$ is also much larger than the neutrino mass-squared differences (this is the so-called ``quasi-degenerate neutrino masses'' regime). Under these circumstances, $m_1^2\simeq m_2^2 \simeq m_3^2\simeq m_l^2$, independent of the mass hierarchy, and $m_{\nu_e}^2\simeq m_l^2\sum_i|U_{ei}|^2=m_l^2$. Hence, the bound $m_l\in[0,2.0]$~eV, quoted in Sec.~\ref{sec:intro}. Incidentally, the muon-type and tau-type neutrino effective masses-squared are given by $m^2_{\nu_{\alpha}}=\sum_im_i^2|U_{\alpha i}|^2$, $\alpha=\mu,\tau$. Given the current bound $m_l<2.0$~eV, both are constrained to be less than $(2.0)^2$~eV$^2$, orders of magnitude smaller than the current kinematical limits obtained from charged-current processes involving muons and taus \cite{pdg}.

\begin{figure}[th]
\centerline{\epsfig{width=0.7\textwidth, file=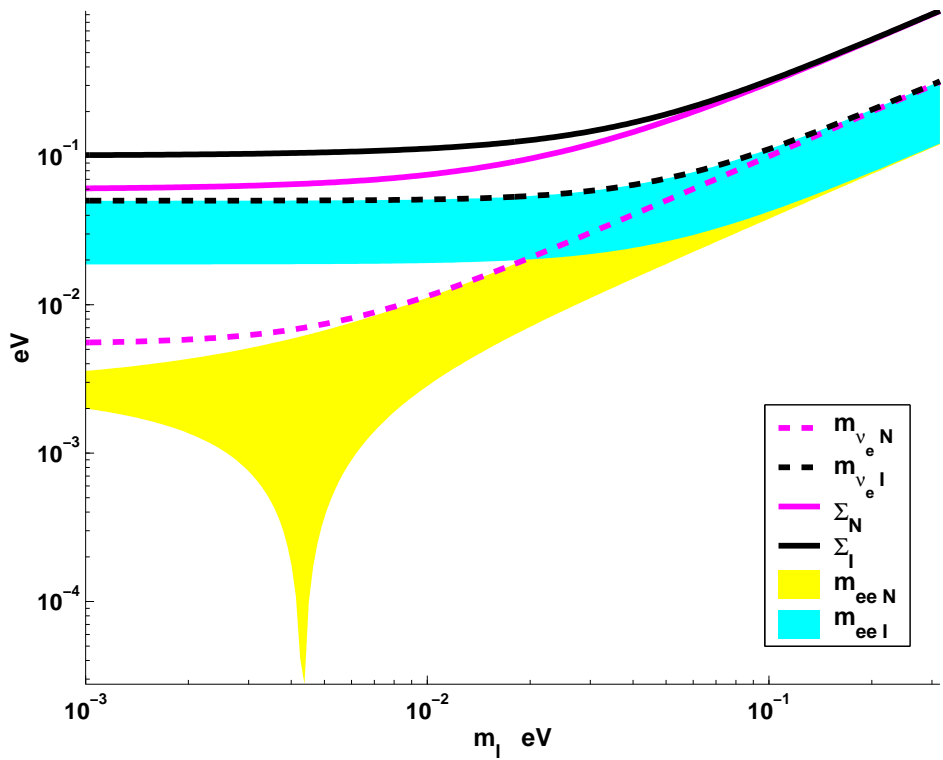}}
\caption{$m_{\nu_e}$ (dotted lines), $m_{ee}$ (solid bands) and $\Sigma$ (solid lines) as a function of the lightest neutrino mass, $m_l$ and the neutrino mass hierarchy (normal $N$ or inverted $I$) for $\Delta m^2_{12}=8.0\times 10^{-5}$~eV$^2$, $\Delta m^{2+}_{13}=2.50\times 10^{-3}$~eV$^2$, $\Delta m^{2-}_{13}=-2.44\times 10^{-3}$~eV$^2$, $\sin^2\theta_{12}=0.31$, and $\sin^2\theta_{13}=0$. See text for details.
\label{fig_ms}}
\end{figure}

In the case $\sin^2\theta_{13}=0$, $m_{\nu_e}$ is simply given by
\begin{eqnarray}
m_{\nu_e}^2&=&m_1^2\cos^2\theta_{12}+m_2^2\sin^2\theta_{12}, \\
&=&m_1^2+\Delta m^2_{12}\sin^2\theta_{12}, \\
&=&m_l^2+\Delta m^2_{12}\sin^2\theta_{12}~~~\rm (normal~hierarchy), \\
&=&m_l^2-\Delta m^2_{13}+\Delta m^2_{12}\sin^2\theta_{12}~~\rm (inverted~hierarchy).
\end{eqnarray}
In the case of a normal mass hierarchy, the expressions above still hold for all values of $m_l$ and nonzero $|U_{e3}|$, as long as $|\Delta m^2_{13}|\sin^2\theta_{13}\ll\Delta m^2_{12}\sin^2\theta_{12}\to \sin^2\theta_{13}\ll 0.01$.  In the case of an inverted hierarchy, the expressions above are valid if $\sin^2\theta_{13}\ll 1$, which is already experimentally guaranteed to be satisfied. Fig.~\ref{fig_ms} depicts $m_{\nu_e}$ as a function of $m_l$ for an inverted and a normal mass hierarchy, and for $\Delta m^2_{12}=8.0\times 10^{-5}$~eV$^2$, $\Delta m^2_{13}=-2.44\times 10^{-3}$~eV$^2$, $\sin^2\theta_{12}=0.31$, and (as already mentioned) $\sin^2\theta_{13}=0$.

We are interested in extracting the neutrino mass hierarchy from experimental information on $m_{\nu_e}$, and address the question in the following way. We assume that all oscillation parameters are measured precisely enough that uncertainties on these are irrelevant (we worry about the precision with which oscillation parameters are measured in the next section). Furthermore, since we are working under the assumption that $|U_{e3}|=0$, we also assume that the neutrino mass hierarchy is not known, {\it i.e.,}\/ there are two values of $\Delta m^2_{13}$ that perfectly fit all neutrino oscillation data. These will be referred to as $\Delta m^{2+}_{13}>0$ and $\Delta m^{2-}_{13}<0$. As discussed in detail in \cite{ours}, it is quite likely that $\Delta m^{2+}_{13}-|\Delta m^{2-}_{13}|\equiv x\neq 0$. The value of $x$ will depend on the details of the measurement of $\Delta m^2_{13}$  \cite{ours}.

Currently, the experimental upper bound on $m_{\nu_e}$ does not allow one to discriminate a normal from an inverted mass hierarchy. This remains true as long as
\begin{equation}
m^2_{\nu_e}>-\Delta m^{2-}_{13}.
\label{eq:mb_bound}
\end{equation}
Indeed, given our current knowledge of $\Delta m^2_{13}$, in order to determine the mass hierarchy from an upper bound (or measurement) of the electron-type effective neutrino mass, one should be able to constrain $m^2_{\nu_e}\lesssim 1.7\times 10^{-3}$~eV$^2$ (the current 99\% upper bound on $\Delta m^{2-}_{13}$ \cite{global_anal}). This can only happen if the mass hierarchy is normal.

In other words, if the neutrino mass hierarchy is inverted, it is not possible to determine the character of the neutrino mass by measuring $m_{\nu_e}$, irrespective of how precise a measurement one is able to perform. Associated to the inverted mass hierarchy measurement of $m_l=m_l^-$, there is a different value of $m_l=m_l^+$ that provides an equally good fit to all neutrino data as long as $m_l^{2+}=m_l^{2-}-\Delta m^{2-}_{13}$ (we remind readers that $\Delta m^{2-}_{13}$ is negative-definite, such that $m_l^+>m_l^-$). If the mass hierarchy is normal and the true value of the lightest mass is $m_l^{2+}>-\Delta m^{2-}_{13}$, there is an equally good fit to all the neutrino data with an inverted hierarchy and a lightest mass $m_l^-$ given by $m_l^{2-}=m_l^{2+}+\Delta m^{2-}_{13}$. If $m_l^{2+}<-\Delta m^{2-}_{13}$, there is no inverted mass hierarchy ``mirror solution.''

In the future, the KATRIN experiment, currently under construction, aims at being sensitive to values of $m_{\nu_e}>0.2$~eV at the 90\% confidence level \cite{KATRIN}. Note that this still satisfies Eq.~(\ref{eq:mb_bound}), such that KATRIN will not be able to establish the neutrino mass hiearchy, even if it observes a nonzero $m_{\nu_e}$ effect.  We are not aware of experimental proposals to significantly improve the sensitivity to $m_{\nu_e}$ beyond the reach of KATRIN.


\subsection{$m_{ee}$ --- Neutrinoless Double Beta Decay}

If the neutrinos are Majorana fermions, lepton number is not a good quantum number.\footnote{To be more precise, $U(1)_{B-L}$ is no longer a global symmetry of the Lagrangian that describes standard model degrees of freedom.} The most powerful probe of lepton number violation is the search for neutrinoless double beta decay, $0\nu\beta\beta$: $^AZ\to^A(Z+2)e^-e^-$. Assuming that the neutrino masses are the only relevant source of lepton number non-conservation, the rate for $0\nu\beta\beta$ is governed by the fundamental process $W^-W^-\to e^-e^-$, mediated by Majorana neutrino exchange.

Schematically,
\begin{equation}
\Gamma_{0\nu\beta\beta}\propto \left|\sum_i U_{ei}^2 \frac{m_i}{Q^2+m_i^2}{\cal M}(m_i^2,Q^2)\right|^2,
\label{eq:mbb}
\end{equation}
where $Q^2$ is a typical energy momentum transfer (squared) associated with the $0\nu\beta\beta$ process --- $Q^2\sim (50)^2$~MeV$^2$ --- and $\cal M$ contains the rest of the kinematics and the nuclear matrix element associated with the $Z\to (Z+2)$ transition. As is well known, the rate for $0\nu\beta\beta$ is proportional to a combination of the neutrino masses that vanishes in the limit $m_i\to 0$, $\forall i$.  This is, of course, expected, since we are assuming that nonzero Majorana neutrino masses are the only source of lepton number violation.

All $m^2_i$ are much smaller than $Q^2$, so we can simplify Eq.~(\ref{eq:mbb}) by expanding the neutrino propagator 
\begin{equation}
\frac{m_i}{Q^2+m_i^2}=\frac{m_i}{Q^2}\left(1-\frac{m_i^2}{Q^2}+O\left(\frac{m_i^4}{Q^4}\right)\right),
\end{equation}
and ${\cal M}(m_i^2,Q^2)={\cal M}(0,Q^2)(1+O(m_i^2/Q^2))$ such that, up to corrections proportional to $m^2_i/Q^2$,
\begin{equation}
\Gamma_{0\nu\beta\beta}\propto |m_{ee}|^2,
\end{equation}
where
\begin{equation}
m_{ee}\equiv \sum_i U_{ei}^2m_i\equiv m_1|U_{e1}|^2e^{i\alpha_1}+m_2|U_{e2}|^2e^{i\alpha_2}+m_3|U_{e3}|^2e^{-2i\delta}
\label{eq:mbb_def}
\end{equation}
 is the effective mass for $0\nu\beta\beta$.\footnote{After electroweak symmetry breaking and in the weak basis where the charged current couplings and the charged lepton mass matrix are diagonal, $m_{ee}$ is the ``$\nu_e\nu_e$'' entry of the neutrino Majorana mass matrix.} We will assume henceforth that $0\nu\beta\beta$ decay experiments are capable of constraining (or measuring) $|m_{ee}|$.

In the case of interest here ($\sin^2\theta_{13}=0$), 
\begin{eqnarray}
|m_{ee}|&=&|m_1\cos^2\theta_{12}e^{i\alpha_1}+m_2\sin^2\theta_{12}e^{i\alpha_2}| \\
&=&\left(m_1^2\cos^4\theta_{12}+m_2^2\sin^4\theta_{12}+2m_1m_2\sin^2\theta_{12}\cos^2\theta_{12}
\cos\alpha\right)^{1/2}, \\
&=&\left(m_l^2\cos^4\theta_{12}+(m_l^2+\Delta m^2_{12})\sin^4\theta_{12}+m_l\sqrt{m_l^2+\Delta m^2_{12}}\frac{\sin^22\theta_{12}}{2}\cos\alpha\right)^{1/2}\rm (normal~hierarchy), 
\end{eqnarray}
and,
\begin{eqnarray}
|m_{ee}|&=&\left((m_l^2-\Delta m^2_{13})\cos^4\theta_{12}+(m_l^2+\Delta m^2_{12}-\Delta m^2_{13})\sin^4\theta_{12} \right. \nonumber \\
&+&\left.\sqrt{(m^2_l-\Delta m^2_{13})(m_l^2+\Delta m^2_{12}-\Delta m^2_{13})}\frac{\sin^22\theta_{12}}{2}\cos\alpha\right)^{1/2}~~~\rm (inverted~hierarchy),
\end{eqnarray}
where $\alpha\equiv \alpha_2-\alpha_1$ is the Majorana phase to which $m_{ee}$ is sensitive (in the limit $\sin^2\theta_{13}=0$). Given the current constraints on $\sin^2\theta_{13}$, the expressions above are an excellent approximation for $m_{ee}$ in the case of an inverted hierarchy, and also apply safely for a normal mass hierarchy as long as $\sqrt{|\Delta m^2_{13}|}\sin^2\theta_{13}\ll\sqrt{\Delta m^2_{12}}\sin^2\theta_{12}\to \sin^2\theta_{13}\ll 0.05$. Fig.~\ref{fig_ms} depicts $|m_{ee}|$ as a function of $m_l$ for an inverted and a normal mass hierarchy, and for $\Delta m^2_{12}=8.0\times 10^{-5}$~eV$^2$, $\Delta m^2_{13}=-2.44\times 10^{-3}$~eV$^2$, $\sin^2\theta_{12}=0.31$, and $\sin^2\theta_{13}=0$. Associated to a fixed value of $m_l$ and a mass hierarchy there is a continuum of  values of $|m_{ee}|$, one for each value of $\cos\alpha\in[-1,1]$.

The minimum (maximal) value of $m_{ee}$ for a fixed value of $m_l$ corresponds to $\cos\alpha=-1~(+1)$. $m_{ee}$ can vanish in the case of a normal mass hierarchy, when $m^2_l\cos^4\theta_{12}=(\Delta m^2_{12}+m_l^2)\sin^4\theta_{12}$. Hence, for
\begin{equation}
m_l^2\sim \Delta m^2_{12}\frac{\sin^4\theta_{12}}{\cos^2\theta_{12}-\sin^2\theta_{12}}\sim (0.004)^2,
\end{equation}
and an inverted mass hierarchy, $m_{ee}$ is especially suppressed.

As in the previous subsection, it is easy to show that if
\begin{equation}
|m_{ee}|>\sqrt{\Delta m^{2-}_{13}}\cos^2\theta_{12}-\sqrt{\Delta m^{2-}_{13}+\Delta m^2_{12}}\sim 0.02~\rm eV,
\end{equation}
a measurement of $|m_{ee}|$ has no discriminatory power when it comes to determining the nature of the neutrino mass hierarchy, irrespective of its precision. As before, such values can only be experimentally ruled out if the mass hierarchy is normal \cite{petcov_pascoli}. Hence, any ``perfect'' fit to all the neutrino data (including a potential measurement of $|m_{ee}|$) obtained under the assumption that the neutrino mass hierarchy is inverted can be matched by an as-perfect fit obtained under the assumption that the mass hierarchy is normal. A very recent and thorough analysis of the capabilities of a measurement of $|m_{ee}|$ to determine the mass hierarchy can be found in \cite{choubey_rodejohann}.

In practice, the situation is significantly more involved, for three main reasons. One is that the extraction of $|m_{ee}|$ from the rate for $0\nu\beta\beta$ is severely clouded by uncertainties in computing the nuclear matrix elements. Different theoretical estimates that make use of different nuclear physics techniques can differ significantly, often by an order of magnitude \cite{nuclear_matrix}. We have nothing to add to the discussion, except for the fact that the situation should improve significantly in the next several years, especially if a positive signal for $0\nu\beta\beta$ is obtained. Another source of confusion is that $\Gamma_{0\nu\beta\beta}$ is only proportional to $|m_{ee}|$ if there are no other ``beyond the standard model'' sources of lepton number violation. We will come back to this issue briefly in the next section, and argue that the combined analysis of  searches for $m_{\nu_e}$, $|m_{ee}|$ and $\Sigma$ (discussed in the next subsection) can shine light on this subject. Finally, the rate for $0\nu\beta\beta$ is only nonzero if lepton number conservation is not exact. As far as this discussion is concerned, this implies that while evidence for $0\nu\beta\beta$ can be translated into a nonzero $|m_{ee}|$, failed attempts to observe  $0\nu\beta\beta$ cannot be translated into an upper bound on $|m_{ee}|$, and hence potentially vital information regarding the nature of the neutrino mass hierarchy. This should be contrasted with failed attempts to measure $m_{\nu_e}$ from tritium beta decay. These can be (much more model-independently) translated into information that may help resolve the nature of the neutrino mass hierarchy. 

Currently, $m_{ee}$ is (conservatively) constrained to be $|m_{ee}|<0.91$~eV at the 99\% confidence level (see \cite{global_anal} and references therein for details), and we will ignore the to-be-confirmed recent evidence for a nonzero rate for $0\nu\beta\beta$ \cite{klapdor}. As in the case for the current bound on $m_{\nu_e}$, no information regarding the neutrino mass hierarchy can be obtained from our current knowledge of $|m_{ee}|$. This would remain true even if we postulated that the neutrinos are Majorana fermions. Near-future experiments with sensitivity to $|m_{ee}|\gtrsim 0.1$~eV are currently being planned, and several of those aim at reaching an ultimate sensitivity to $|m_{ee}|\gtrsim 0.01$~eV \cite{0nubb_future}. It is fair to say that significant R\&D efforts (plus time and resources) are necessary in order to improve the sensitivity to $|m_{ee}|$ beyond that.

\subsection{$\Sigma$ --- Cosmological Observables}

There is a significant amount of indirect evidence that the Universe is filled with a very cold thermal population of relic neutrinos --- the cosmic neutrino background. While we have no direct evidence that these really exist, their presence is required in order to match predictions of big bang nucleosynthesis with measurements of the primordial abundances of light nuclei \cite{BBN}. Indeed, before LEP experiments measured the invisible $Z$-boson width, bounds from big bang nucleosynthesis were known to provide the most stringent constraints on the number of light neutrino species \cite{pdg}.

If neutrinos have mass, the cosmic neutrino background contributes to the energy budget of the Universe as hot dark matter. According to the very successful (if not very palatable) concordance cosmological model \cite{recent_cosmology}, the fraction of hot dark matter in the Universe is constrained to be small. Given our current understanding of neutrino masses, $\Omega_{\rm hot}\simeq\sum m_i/(48~\rm eV)$, such that an upper bound on $\Omega_{\rm hot}$ can be translated directly into an upper bound on
$\Sigma\equiv \sum_i m_i$.
As was the case in previous subsections, cosmological observables are potentially sensitive to more than the sum of the neutrino masses, at least in principle. For example, hot dark matter has the property of preventing matter from ``clumping'' in the early Universe, leading to a suppression of power at small scales in the matter distribution power spectrum \cite{mc_bound_review}. Potential ``kinks'' in the spectrum are related to the mass of the individual hot dark matter species. We will ignore this possibility and will assume that cosmological observables are only sensitive to $\Sigma$. 

$\Sigma$ depends only on the neutrino masses, and can be expressed in terms of $m_l$ and the mass-squared differences:
\begin{eqnarray}
\Sigma&=&m_1+\sqrt{m_1^2+\Delta m^2_{12}}+\sqrt{m_1^2+\Delta m^2_{13}}, \label{eq:mc_def}\\
&=&m_l+\sqrt{m_l^2+\Delta m^2_{12}}+\sqrt{m_l^2+\Delta m^2_{13}}~~~(\rm normal~hierarchy), \\
&=&m_l+\sqrt{m_l^2-\Delta m^2_{13}}+\sqrt{m_l^2+\Delta m^2_{12}-\Delta m^2_{13}}~~~(\rm inverted~hierarchy).
\end{eqnarray}
$\Sigma$ is the only observable discussed in this section that, in the limit $\theta_{13}\to 0$, is sensitive to $\Delta m^{2+}_{13}$, the value of the atmospheric mass hierarchy obtained assuming that the mass hierarchy is normal. Fig.~\ref{fig_ms} depicts $\Sigma$ as a function of the lightest neutrino mass, $m_l$ for $\Delta m^2_{12}=8.0\times 10^{-5}$~eV$^2$, and $\Delta m^{2+}_{13}=2.50\times 10^{-3}$~eV$^2$, $\Delta m^{2-}_{13}=-2.44\times 10^{-3}$~eV$^2$, for both neutrino mass hierarchies.

As before, if
\begin{equation}
\Sigma>\sqrt{-\Delta m^{2-}_{13}}+\sqrt{\Delta m^2_{12}-\Delta m^{2-}_{13}},
\label{mc_bound}
\end{equation}
a measurement of $\Sigma$ is not sensitive to the nature of the neutrino mass hierarchy, independent of its precision. In light of the discussion in the two previous subsections, it should not be surprising that Eq.~(\ref{mc_bound}) is automatically satisfied in the case of an inverted mass hierarchy, but could be violated in the case of a normal one, as long as $m_l$ is small enough ($m_l\lesssim 0.02$~eV). Quantitatively, for every candidate value of the lightest mass $m_l=m_l^-$ obtained by postulating that the mass hierarchy is inverted, there is another lightest mass value $m_l=m_l^+$ that fits the neutrino data just as well as long as one postulates that the mass hierarchy is normal and $m_l^+$ satisfies 
\begin{equation}
m_l^++\sqrt{m_l^{2+}+\Delta m^2_{12}}+\sqrt{m_l^{2+}+\Delta m^{2+}_{13}}=m^-_l+\sqrt{m_l^{2-}-\Delta m^{2-}_{13}}+\sqrt{m_l^{2-}+\Delta m^{2}_{12}-\Delta m^{2-}_{13}}.
\end{equation}

As with measurements of $|m_{ee}|$, the map between cosmological measurements and $\Sigma$ is nontrivial and model dependent. The cosmological determination of $\Sigma$ depends on the history of the Universe and its particle content, and physics beyond the standard model and/or concordance cosmology can easily obscure, enhance or completely erase any relationship between cosmological observables and $\Sigma$ \cite{mc_bound_review,avoid_mc}.

According to \cite{global_anal}, combined analysis of data from WMAP, SDSS and Lyman-alpha forest surveys constrain $\Sigma<0.94$~eV at the 99\% confidence level. Different analyses of the same data or subsets thereof quote similar bounds (within 50\%) \cite{mc_bound_review}. Upcoming missions have the potential to significantly improve on the current sensitivity to a nonzero value of $\Sigma$. 
It is expected that future data on the cosmic microwave background should be sensitive to $\Sigma\gtrsim 0.1$~eV \cite{sigma_shift}, while current studies of the capabilities of weak lensing probes seem to indicate that these might be sensitive to $\Sigma$ values below the 0.1~eV level \cite{lensing}. For an overview see, for example, \cite{sigma_shift}. It has recently been speculated that one may indeed reach $\Sigma$ values as low as 0.03~eV \cite{wang_etal}.

\setcounter{footnote}{0}
\setcounter{equation}{0}
\section{Combined Fits to the Mass Hierarchy}
\label{sec:fits}

It has been widely recognized in the literature \cite{many,petcov_wolfenstein,lisi_silk,new_petcov} that the combined analysis of searches for $m_{\nu_e}$, $m_{ee}$,\footnote{Henceforth, to simplify the notation, we define $m_{ee}\equiv|m_{ee}|$.} and/or $\Sigma$ provide qualitatively more information than the separate analysis of each observable, including potential information regarding the mass hierarchy \cite{petcov_wolfenstein}. For example, a lower bound on $m_{\nu_e}$ combined with an upper bound for $m_{ee}$ may help determine, with the addition of some reasonable assumptions, that the neutrinos are Dirac fermions, while a large $m_{\nu_e}$ value, combined with an upper bound on $\Sigma$ would lead one to conclude that there is more to our understanding of particle physics or the history of the Universe than currently accepted. Here, we discuss the circumstances under which measurements of (or upper bounds for) $m_{\nu_e}$, $m_{ee}$, and/or $\Sigma$ help us determine the mass hierarchy.

It is easy to understand why a combined analysis should prove more powerful. The value of $m_l^+$ that renders a normal hierarchy fit to, say $\Sigma$, compatible with an inverted hierarchy one (associated to a value $m_l^-$ of the lightest mass) need not agree with the the equivalent quantity obtained from analyzing measurements of $m_{ee}$ or $m_{\nu_e}$. For example, if the true value of $m_l$ vanishes and the mass hierarchy is inverted, the mirror value $m_l=m_l^+$ that renders $\Sigma$ ``the same'' if one assumes that the mass hierarchy is normal is $m_l^+(\Sigma)\sim 0.5\sqrt{\Delta m^{2-}_{13}}\sim 0.02$~eV. On the other hand, the equivalent for $m_{\nu_e}$ is $m_l^+(m_{\nu_e})=\sqrt{\Delta m^{2-}_{13}}\sim0.05$~eV. Hence, for the wrong-hierarchy hypothesis, $m_l(m_{\nu_e})\neq m_l(\Sigma)\neq m_l(m_{ee})$ --- even if the true neutrino mass hierarchy is inverted. This indicates that combined analyses of the three observables discussed in Sec.~\ref{sec:probes} can improve, sometimes significantly, our ability to determine the neutrino mass ordering.

\begin{figure}[th]
\begin{center}
\includegraphics[scale=0.7]{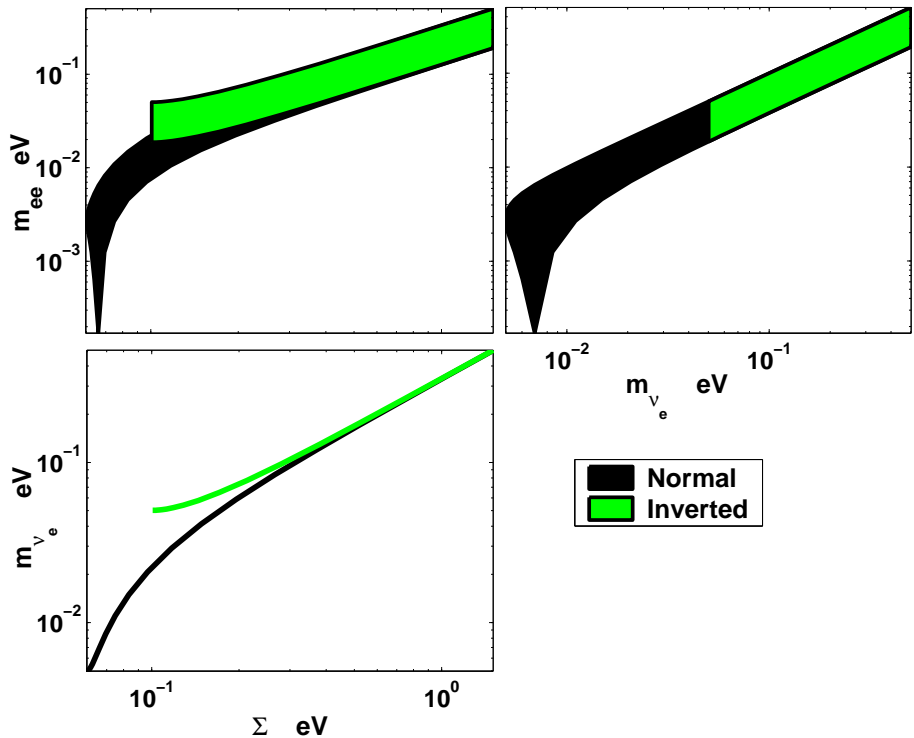}
\end{center}
\caption{Values of $m_{\nu_e}$, $|m_{ee}|$, and $\Sigma$
for both hierarchies, for $\Delta m^2_{12}=8.0\times 10^{-5}$~eV$^2$,
$\Delta m^{2+}_{13}=2.50\times 10^{-3}$~eV$^2$,  $\Delta m^{2-}_{13}=-2.44\times 10^{-3}$~eV$^2$,
$\sin^2\theta_{12}=0.31$ and $\sin^2\theta_{13}=0$. The value of $m_l$ varies between 0 and 0.5~eV and the relevant  Majorana phase spans its entire allowed physical range: $\cos\alpha\in[-1,1]$. See text for details.}\label{figure:2by2}
\end{figure}

Figure~\ref{figure:2by2} depicts the effective masses pairwise
for both hierarchies, assuming $\Delta m^2_{12}=8.0\times 10^{-5}$~eV$^2$,
$\Delta m^{2+}_{13}=2.50\times 10^{-3}$~eV$^2$,  $\Delta m^{2-}_{13}=-2.44\times 10^{-3}$~eV$^2$,
$\sin^2\theta_{12}=0.31$ and $\sin^2\theta_{13}=0$, allowing $m_l$ to vary between 0 and 0.5~eV, and the relevant  Majorana phase to span its entire allowed physical range: $\cos\alpha\in[-1,1]$.  
One can easily identify sets of potential 
measurements that, with sufficiently small uncertainties, would
imply a specific mass hierarchy.\footnote{The usefulness of looking at the non-oscillation observables as depicted in Fig.~\ref{figure:2by2} was pointed out in \cite{lisi_silk}.}  In particular, it is easy to
verify, and improve on, the claims made in section \ref{sec:probes}
regarding the implication of the normal scheme for sufficiently
small measurements of the effective masses.  Evidence for the inverted hierarchy can also arise from
such naive considerations.  For instance, by examining the upper
left panel it is clear that the inverted hierarchy could be
established if accurate measurements of $m_{ee}$ and $\Sigma$ were
made near 0.04 eV and 0.1 eV, respectively.  The same conclusion is
obtained if only an upper bound exists for $\Sigma$ near $0.2$~eV,
while a measurement sets $m_{ee}$ near $0.05$~eV. It is easy to see that combined measurements of $m_{\nu_e}$ and $m_{ee}$ cannot distinguish an inverted mass hierarchy from a normal, quasi-degenerate scheme --- knowledge of $\Sigma$ is extremely valuable when it comes to establishing an inverted mass hierarchy. 

This simple picture is smeared out when realistic conditions are
taken into account.  The inclusion of parameter uncertainties will
naturally wash out some of the interesting regions in this
measurement space, but most of the main features remain robust.  A
complete statistical analysis, quantifying the very visual and
intuitive line of reasoning outlined here, is presented below.

\subsection{Parameters and Data Analysis}

In order to quantitatively compare the different neutrino mass hierarchies we proceed as follows. We assume that all oscillation parameters $\Theta_i=\Delta m^2_{13},\Delta m^2_{12},\sin^2\theta_{12},\sin^2\theta_{13}$ are measured to be $\overline{\Theta_i}\pm\sigma_{\Theta_i}$, while the three non-oscillation parameters are constrained to be $m_{\nu_e}=\overline{m_{\nu_e}}\pm\sigma_{\nu_e}$, $m_{ee}=\overline{m_{ee}}\pm\sigma_{ee}$, and $\Sigma=\overline{\Sigma}\pm\sigma_{\Sigma}$. Note that we are assuming that the uncertainty related to the computation of the $0\nu\beta\beta$ nuclear matrix element is included in $\sigma_{ee}$. We assume no information concerning the three CP-odd phases $\alpha_1,\alpha_2,\delta$. We refer to all these measurements as the data. We further postulate that the neutrinos are Majorana fermions, and that there are no other sources of lepton number violation. Finally, we assume concordance cosmology and ``standard'' particle physics.

Armed with these results, we construct a $\chi^2$ function defined by
\begin{equation}
\chi^2(m_l,\alpha_1,\alpha_2,\delta,\Theta) = \frac{[\overline{m_{\nu_e}} -
m_{\nu_e}(m_l,\Theta)]^2}{\sigma_{\nu_e}^2}
+\frac{[\overline{m_{ee}} -
m_{ee}(m_l,\Theta,\alpha_1,\alpha_2,\delta)]^2}{\sigma_{ee}^2} +
\frac{[\overline{\Sigma} - \Sigma(m_l,\Theta)]^2}{\sigma_{\Sigma}^2}
+ \sum_i{\frac{[\overline{\Theta_i} -
\Theta_i]^2}{\sigma_{\Theta_i}^2}}.
\end{equation}
This function also depends implicitly on the choice of
hierarchy from the structure of the effective masses
$m_{\nu_e}$, $m_{ee}$, and $\Sigma$ given, respectively, by Eqs.~(\ref{eq:mb_def},\ref{eq:mbb_def},\ref{eq:mc_def}).

For each hypothesis concerning the mass hierarchy, we minimize $\chi^2$ with respect to all neutrino parameters ($\Theta_i$, $m_l$, and the two relevant CP-odd phases [say, $\alpha_2-\alpha_1$, and $-2\delta-\alpha_1$]), and establish if a good fit to the data can be obtained. This is done by comparing the minimum value of $\chi^2=\chi^2_{\rm min}$ with expectations for three degrees of freedom. In particular, $\chi^2_{\rm min} < 7.8$ implies that the theory fits the data at the 95\% confidence level. If only one of the hypotheses concerning the mass hierarchy fits the data at the 95\% confidence level (or better), we conclude that it is established (at least) at the 95\%, while if both hypotheses prove to fit the data at the 95\% confidence, we conclude that the mass-hierarchy cannot be determined. Finally, if neither hypothesis fits the data, we are forced to conclude that there is a flaw in our theoretical model. Candidate flaws include the hypothesis that the neutrino is a Majorana fermion, that concordance cosmology is correct, etc.

The different data sets used in our numerical
analysis are listed in Table~\ref{table:parameters}.  The best fit
values \cite{global_anal} from the global analysis of neutrino
oscillations were used for $\overline{\Theta}$, except for $\overline{\Delta m^2_{13}}$. In the case of a normal mass hierarchy, we pick $\Delta m^{2+}_{13}=2.50\times 10^{-3}$~eV$^2$, while in the case of a normal mass hierarchy, we choose $\Delta m^{2-}_{13}=-\Delta m^{2+}_{13}+x$, where $x=2\Delta m^2_{12}\cos2\theta_{12}$, in order to take into account that fact that, in the case of very precise measurements of $\Delta m^2_{13}$, different hypothesis regarding the mass hierarchy yield different values for $|\Delta m^2_{13}|$ \cite{ours}. The value of $x$ above agrees with what one expects to obtain in accelerator $\nu_{\mu}\to\nu_{\mu}$ studies at relatively short baselines ($L\lesssim 3000$~km) or relatively large energies ($E_{\nu}\gtrsim 1$~GeV). We refer readers to \cite{ours} for details.
\begin{table}
\centering 
\caption{Input values, including uncertaintites, for the simulated data sets. `x-axis' and `y-axis' refer to Fig.~\ref{figure:run}. The `comments' column refers to the experiment (or class of experiments) expected to contribute most significantly to the uncertainty with which each neutrino observable is assumed to be measured.}
\begin{tabular}{c|c c|c c|c c|c} 
\hline
 & Run 1 & & Run 2 & & Run 3 &~ \\
Parameter & Central & $\sigma$ & Central & $\sigma$ & Central & $\sigma$ & comments  \\ [0.5ex] 
\hline\hline
$\Delta m^2_{12} (10^{-5}~{\rm eV}^2)$  & 8.0 & 5\% & 8.0 & 5\% & 8.0 & 5\% & Reactor \cite{solar}\\ 
$\Delta m^{2+}_{13} (10^{-3}~{\rm eV}^2)$  & 2.5 & 1\% & 2.5 & 1\% & 2.5 & 1\% & T2K, NO$\nu$A, MINOS \cite{Lindner} \\
$\Delta m^{2-}_{13} (10^{-3}~{\rm eV}^2)$  & --2.44 & 1\%& --2.44 & 1\% & 2.5 & 1\% & T2K, NO$\nu$A, MINOS \cite{Lindner} \\
$\sin^2\theta_{12}$  & 0.31 & 5\% & 0.31 & 5\% & 0.31 & 5\% & Reactor \cite{theta12}\cite{solar} \\
$\sin^2\theta_{13}$  & 0 & 0.01 & 0 & 0.01 & 0  & 0.01  & T2K, NO$\nu$A,D-CHOOZ \cite{Lindner}\\
\hline
$m_{\nu_e} ~{\rm(eV)}$  & 0 & 0.1 & 0 & 0.1 & 0 & 0.1 & KATRIN \cite{KATRIN} \\
$m_{ee}~{\rm (eV)}$ & y-axis & 0.01 & y-axis & 0.05 & y-axis & 0.1 & \\
$\Sigma~{\rm (eV)}$  & x-axis & 0.01 & x-axis & 0.05 & x-axis & 0.1 & \cite{mc_bound_review} \\ [1ex] 
\hline 
\label{table:parameters}
\end{tabular}
\end{table}

In order to anticipate future developments, we assume that the  uncertainties on the various oscillation parameters are smaller than the current ones. For instance, the uncertainty on the
atmospheric mass-squared difference $|\Delta m^2_{13}|$ should be
reduced to $\sim 3\%$ in the future, due to potential results from  off-axis beam experiments,
such as T2K and NO$\nu$A \cite{Lindner}, and future endeavors can improve on that \cite{BNL}. Similarly, we expect that the  the solar parameters, $\Delta m^2_{12}$ and $\sin^2\theta_{12}$ can be measured at the 5\% level in, say, precise measurements of the low energy solar neutrino flux \cite{low_solar}, improved reactor
experiments boasting greater statistics and reduced systematic
errors conducted at baselines at (or near) the first survival
probability minimum \cite{theta12,solar}, or with broadband long-baseline $\nu_{\mu}\to\nu_e$ studies \cite{BNL}.

\subsection{Results}

\begin{figure}[th]
\begin{center}
\includegraphics[scale=0.7]{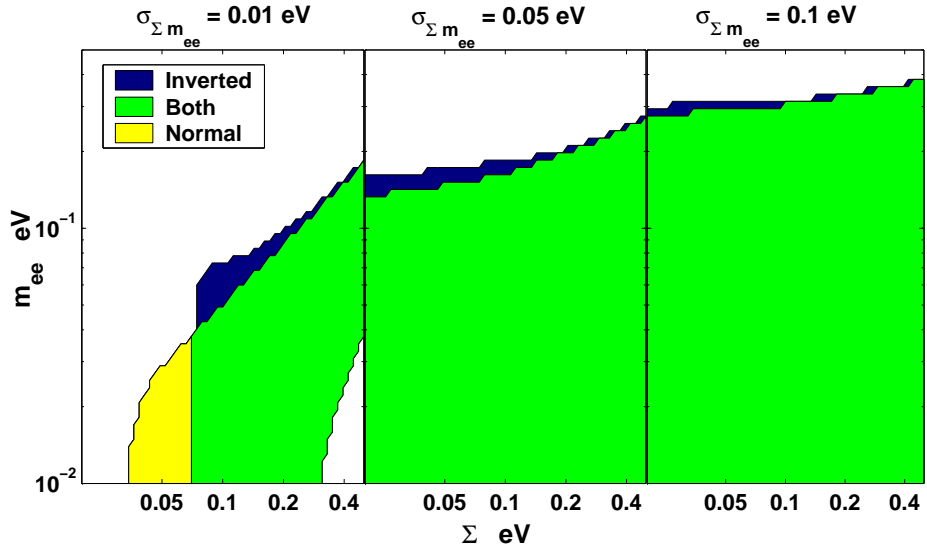}
\end{center}
\caption{Regions of the $\overline{\Sigma}\times\overline{m_{ee}}$ plane consistent with both a normal and an inverted neutrino mass hierarchy (green [gray]), consistent only with a normal mass hierarchy (yellow [lightest gray]), consistent only with an inverted mass hierarchy (blue [dark]), or inconsistent with either hypothesis concerning the mass hierarchy (white). Input parameters and expected uncertainties are tabulated in Table~\ref{table:parameters}. LEFT: $\sigma_{m_{ee}}=\sigma_{\Sigma}=0.01$~eV, MIDDLE:  $\sigma_{m_{ee}}=\sigma_{\Sigma}=0.05$~eV, and RIGHT:  $\sigma_{m_{ee}}=\sigma_{\Sigma}=0.1$~eV. See text for details.}\label{figure:run}
\end{figure}

Figure~\ref{figure:run} depicts regions of the $\overline{\Sigma}\times\overline{m_{ee}}$ plane consistent with both a normal and an inverted neutrino mass hierarchy (green [gray]), consistent only with a normal mass hierarchy (yellow [lightest gray]), consistent only with an inverted mass hierarchy (blue [dark]), or inconsistent with either hypothesis concerning the mass hierarchy (white). The uncertainties with which $\Sigma$ and $m_{ee}$ are expected to be measured, $\sigma_{m_{ee}}$ and $\sigma_{\Sigma}$, are held fixed for all values of $\overline{\Sigma}$, $\overline{m_{ee}}$, and are both equal to 0.01~eV in ``Run 1'' (left panel), 0.05~eV in ``Run 2'' (middle panel) and 0.1~eV in ``Run 3'' (right panel). Throughout, we assume that $m_{\nu_e}$ is experimentally constrained to be less than 0.1~eV (at the one sigma level).\footnote{This is in rough agreement with the sensitivity of the KATRIN experiment, which can exclude $m_{\nu_e}<0.2$~eV at the 90\% confidence level if it does not observe any evidence for nonzero neutrino masses \cite{KATRIN}.} In the absence of new sources of lepton number violation, this upper bound on $m_{\nu_e}$ implies $|m_{ee}|\lesssim 0.1$~eV and, in the absence of nonstandard cosmology, $\Sigma\lesssim 0.3$~eV. We concentrate our exploration of the the parameter space to this region. If, on the other hand, the KATRIN experiment obtained statistically significant evidence that $m_{\nu_e}$ is nonzero (say, $m_{\nu_e}=0.3\pm 0.1$~eV) we would be constrained to the quasi-degenerate neutrino mass scenario, in which it is virtually impossible to determine the mass hierarchy by precisely measuring $m_{\nu_e}$, $m_{ee}$, and $\Sigma$ (see Fig.~\ref{figure:2by2}).

If both $\Sigma$ and $m_{ee}$ are only poorly measured ($\sigma_{m_{ee}}=\sigma_{\Sigma}=0.1$~eV,  Fig.~\ref{figure:run}(RIGHT)) it is, in general, not possible to determine the mass hierarchy, irrespective of $\overline{\Sigma}$ and $\overline{m_{ee}}$. The only exception corresponds to very large values of $\overline{m_{ee}}$, where, given an upper bound on $\Sigma$, only an inverted mass hierarchy provides a good fit. Very large values of $\overline{\Sigma}$ and $\overline{m_{ee}}$ are inconsistent with the upper bound on $m_{\nu_e}$, and, if observed, need to be interpreted as evidence for new cosmology/particle physics and sources of lepton number nonconservation other than the neutrino masses. If $\Sigma$ and $m_{ee}$ are measured somewhat more precisely ($\sigma_{m_{ee}}=\sigma_{\Sigma}=0.05$~eV, Fig.~\ref{figure:run}(MIDDLE)), the situation is qualitatively the same: one will not be able to determine the mass hierarchy from non-oscillation probes of neutrino masses unless $m_{ee}$ is measured to be large ($\overline{m_{ee}}\gtrsim 0.1$~eV) and $\Sigma$ is ``small'' ($\overline{\Sigma}\lesssim 0.1$~eV), in which case only the inverted mass hierarchy provides a good fit to the data. 

Only when both $\Sigma$ and $m_{ee}$ are measured very precisely ($\sigma_{m_{ee}}=\sigma_{\Sigma}=0.01$~eV, Fig.~\ref{figure:run}(LEFT)) do we start to also positively discriminate a normal mass hierarchy. If $\overline{\Sigma}\lesssim 0.07$~eV, only the normal mass hierarchy hypothesis is capable of fitting the data, and only if $\overline{m}_{ee}\lesssim 0.04$~eV. As before, large $\overline{m_{ee}}$ values, appropriately correlated with large $\overline{\Sigma}$ values, point to an inverted mass hierarchy. Furthermore, when the sensitivity of both probes reaches the 0.01~eV level, we fail to fit the data if $\overline{m}_{ee}\lesssim 0.03$~eV and $\overline{\Sigma}\gtrsim0.3$~eV. If faced with such a scenario, we would be tempted to conclude that the neutrinos are Dirac fermions. Finally, note that, if $\sigma_{\Sigma}\simeq 0.01$~eV, we are guaranteed, in the absence of new cosmology/particle phisics, to obtain a nonzero value for $\overline{\Sigma}$ --- hence the region in Fig.~\ref{figure:run}(RIGHT) corresponding to $\overline{\Sigma}\lesssim 0.02$~eV, for all $\overline{m_{ee}}$, is white. 

As alluded to above, if $m_{ee}\gtrsim 0.1$~eV {\sl and} $\Sigma\gtrsim 0.3$, we enter the quasi-degenerate neutrino mass regime, where discrimination between the two mass orderings is virtually unachievable. This behavior can be clearly observed in all panels of Fig.~\ref{figure:run}.

\setcounter{footnote}{0}
\setcounter{equation}{0}
\section{Concluding Remarks}
\label{sec:end}

While we have discovered that neutrinos have mass, there are still qualitative aspects of the neutrino mass spectrum that remain unknown. We have been able to measure, with good precision, the (absolute value of the) two mass-squared differences, but know very little about the magnitude of the individual neutrino masses. We are also ignorant when it comes to the neutrino mass hierarchy. We don't know if the neutrino masses are ``normal ordered'' --- $m_1^2<m_2^2<m_3^2$ --- or whether the mass hierarchy is ``inverted'' --- $m_3^2<m_1^2<m_2^2$ (such that $\Delta m^2_{12}\ll m_2^2,m_1^2$). The reason we have failed (so far) to discover the neutrino mass hierarchy via neutrino oscillations is that both $|U_{e3}|^2$ and $\Delta m^2_{12}/|\Delta m^2_{13}|$ happen to be small.

Next-generation, long-baseline neutrino oscillation experiments are poised to uncover the ordering of the neutrino masses, but they can only be successful if $|U_{e3}|^2$ is large enough. If, on the other hand, $|U_{e3}|^2$ is too small, it is quite challenging to determine the neutrino mass hierarchy via neutrino oscillations \cite{ours}. Other probes are necessary.
Here we have studied in detail whether combined information from future studies of neutrinoless double beta decay ($m_{ee}$), tritium beta-decay ($m_{\nu_e}$) and cosmological probes of the energy composition of the Universe ($\Sigma$) can help establish the neutrino mass hierarchy. 

Our results are summarized in Fig.~\ref{figure:run} and, as expected, the situation is quite challenging. Assuming we can rule out (at the one sigma level) $m_{\nu_e}>0.1$~eV and probe $m_{ee}$ and $\Sigma$ with $\sigma_{m_{ee}}\sim\sigma_{\Sigma}\sim 0.05$~eV, we can only hope to determine the neutrino mass hierarchy from nonoscillation experiments if it is inverted {\sl and} if the Majorana phase to which $m_{ee}$ is sensitive is close to $0$ (``constructive interference,'' $\cos\alpha\sim 1$).\footnote{We remind readers that we are always assuming the value of $|U_{e3}|$ to be vanishingly small.} If $\sigma_{m_{ee}}\sim\sigma_{\Sigma}\sim 0.01$~eV, the situation is significantly improved, and there is also the possibility of establishing a normal mass hierarchy if $\Sigma\lesssim 0.07$~eV and $m_{ee}\lesssim 0.04$~eV. In the next several years ($\lesssim 10$), it is reasonable to expect that we will be somewhere between Fig.~\ref{figure:run}(RIGHT) and Fig.~\ref{figure:run}(MIDDLE), while it is not overly optimistic to assume we will eventually reach somewhere between Fig.~\ref{figure:run}(MIDDLE) and Fig.~\ref{figure:run}(LEFT).

It is curious to note that the ``main goal'' of non-oscillation neutrino experiments is, arguably, {\sl not} to determine the mass hierarchy. These experiments are usually associated with establishing if the neutrinos are Majorana fermions, measuring the lightest neutrino mass $m_l$, and, ultimately, probing whether the Majorana phases are nontrivial ($\alpha_1,\alpha_2\neq 0,\pi$). While we have not concentrated on these issues here, it is clear that in the green [gray] regions of Fig.~\ref{figure:run}, one is not able to properly measure $m_l$, even when $m_{ee}$ and $\Sigma$ are measured to be nonzero with good precision. In those regions, for every normal-hierarchy measurement of $m_l=m_l^+\pm\delta^+_{m_l}$, there is an inverted hierarchy $m_l=m_l^-\pm\delta^-_{m_l}$ that fits the data just as well.

We reemphasize that, in order to convert measurements of $\Gamma_{0\nu\beta\beta}$ and cosmological data into information regarding neutrino masses, we are required to rely on several untested hypothesis, including the fact that Majorana active neutrino masses are the only source of lepton number violation and the fact that we quantitatively understand the history of the Universe from times slightly before big-bang nucleosynthesis until the formation of the observed large-scale structure. It is reassuring that combined measurements of $m_{\nu_e}$, $m_{ee}$, and $\Sigma$ can help falsify these hypotheses in some regions of the parameter space, but there is a finite possibility that, after many years of data analysis, we will be left in a rather ``confused'' state. This would happen, for example, if the mass hierarchy were normal, $m_{l}\lesssim 0.01$~eV and $\sigma_{\Sigma}\gtrsim 0.03$~eV.

We conclude by pointing out that, as far as establishing the neutrino mass ordering is concerned, we would profit tremendously with more precise information on $m_{\nu_e}$. A measurement of (or upper bound for) $m_{\nu_e}$ at the several~$\times 10^{-2}$~eV level, combined with $\sigma_{\Sigma}$ slightly below the 0.1~eV level would be sensitive to the mass hierarchy (see Fig.~\ref{figure:2by2}), and it is easy to appreciate that extracted values of $m_{\nu_e}$ are significantly less model dependent than those of $m_{ee}$ and $\Sigma$.

{\bf Note Added:} While this work was being completed, \cite{new_petcov} was posted in the preprint ArXiv's. It shares several of the results presented here (with a slightly different treatment of the ``data''), and contains a very detailed discussion of the capabilities of $\Sigma$ and $m_{ee}$ (including a detailed treatment of the uncertainties related to nuclear matrix elements) to establish the neutrino mass hierarchy and the existence of nontrivial Majorana phases, and to measure $m_l$. Our contribution, on the other hand, contains a more detailed analysis (including a detailed qualitative discussion) of the capabilities of combined $m_{\nu_e}$, $m_{ee}$, and $\Sigma$ measurements to uncover the neutrino mass ordering, motivated by the fact that it will remain unknown if $|U_{e3}|$ is vanishingly small, and also discusses the importance of measuring (or further constraining) $m_{\nu_e}$.

\section*{Acknowledgments}

We are happy to thank Boris Kayser for enlightening conversations and comments on the manuscript,  Thomas Schwetz for comments on the manuscript, and Silvia Pascoli for words of encouragement. This work is sponsored in part by the US Department of Energy Contract DE-FG02-91ER40684.

 \end{document}